\documentclass[11pt,twoside]{article}
\usepackage{amsmath,amsfonts,amssymb}
\usepackage{latexsym}
\usepackage{dcolumn}
\usepackage{graphicx,epsfig}
\usepackage{amsthm}
\usepackage{amsmath}
\usepackage{cite}
\usepackage{color}

\begin{document}

\setcounter{page}{1}

\pagestyle{plain} \vspace{1 cm}

\begin{center}
{\large{\bf {Thermodynamics of Horndeski Black Holes with Generalized Uncertainty Principle}}}\\

\vspace{1 cm}
{\bf Mohaddeseh Seifi$^{\dag,}$}\footnote{m.seifi@stu.umz.ac.ir}\quad,\quad {\bf Akram S. Sefiedgar$^{\dag,}$}\footnote{a.sefiedgar@umz.ac.ir}    \\
\vspace{0.5 cm}
$^{\dag}$Department of Theoretical Physics, Faculty of Basic Sciences,\\ University of Mazandaran,\\
P. O. Box 47416-95447, Babolsar, Iran\\

\end{center}

\vspace{1.5cm}

\begin{abstract}
Horndeski theory is the most general scalar-tensor extension of General Relativity with second order field equations. It may be interesting to study the effects of the Generalized Uncertainty Principle on a static and asymptotically flat shift symmetric solutions of the Horndeski black holes.
With this motivation, here we obtain the modified black hole temperatures in shift symmetric Horndeski gravity by employing the Generalized Uncertainty Principle. Using the corrected temperature, the entropy and heat capacity are calculated with details. We also investigate the tunneling probability of particles from Horndeski black holes horizon and possible correlations between the emitted modes (particles).
\\
\textbf{PACS:} 04.20.Dw, 04.50.Kd, 04.70.-s, 04.70.Dy\\
\textbf{Key Words:} Generalized Uncertainty Principle, Quantum Gravity, Horndeski Theory, Black Hole thermodynamics.
\end{abstract}

\vspace{2cm}
\newpage
\section{Introduction}
At present, it is believed that Nature can be described by quantum mechanics and general relativity. In 1915, Albert Einstein proposed General Relativity (GR) that is able to successfully describe physical phenomena in astrophysics and cosmology \cite{Berti2015, Clifford2014}. Besides all the significant achievements, general relativity can not describe some theoretical and observational issues. One of the known failures of GR is the problem of black hole singularities that leads to various black hole spacetimes singularities. It seems that GR is cursed with own solutions, i.e., black holes. Also, the cosmological constant problem and the issue of the dark matter/energy can not be explained by GR (for more details see for instance\cite{Riess1998, Perlmutter1999, Sofue2001}). During the recent decades, many efforts have been undertaken to construct a more comprehensive theory. One way to modify GR is reconstructing the geometric part of the Einstein field equations \cite{Clifton2012, Papantonopoulos2015}. A special class, that is, the most general scalar-tensor theory with second order field equations, proposed by Horndeski in 1970s \cite{Horndeski1974}. Recently, Hondeski theory has been received much attention and investigated in astrophysics and cosmology \cite{Saridakis2010, Charmousis2012, Zumalac2014, Gleyzes2015, Langlois2016, Kobayashi2019}. More attractively, Horndeski black holes have been investigated. For instance, spherically symmetric and static solutions \cite{Rinaldi2012, Babichev2016, Babichev2017}, black hole solutions in the presence of a cosmological constant and magnetic field \cite{Anabalon2014, Cisterna2014}, the observational results and gravitational lensing effects for Horndeski black holes \cite{Badia2017, Kumar2022*, Afrin2022} are studied. Moreover, thermodynamics of Horndeski black holes are studied, Hawking temperature and entropy and circular orbits are investigated also in Refs. \cite{Miao2016, Hajian2021, Walia2022, Salahshoor2018}.

Trying to construct a quantum theory of gravity leads to a minimal measurable length of the order of the Planck length, $\ell_P \sim 10^{-35} m $. Most quantum gravity approaches such as string theory \cite{Veneziano1986, Amati1989, Gross1987, Konishi1990}, loop quantum gravity \cite{Garay1995} and quantum geometry \cite{Capozziello2000} predict the existence of a minimal measurable length in spacetime \cite{Maggiore1993}. Also, the existence of a minimal measurable length can be supported from micro-black hole Gedanken experiment \cite{Scardigli1999}. In GUP concept, the Heisenberg Uncertainty Principle (HUP) is modified to the so called Generalized Uncertainty Principle (GUP) \cite{Kempf1994, Kempf1995, Hinrichsen1996, Kempf1997, Hossenfelder2012, Nozari2012, Hossenfelder2013, Tawfik2014}. Incorporation of GUP effects in standard quantum mechanics' problems reveals several novel corrections and modifies the results in high energy regime \cite{Nozari2006, Harbach2006, Das2008, Basilakos2010, Ali2011}. On the other hand, Doubly Special Relativity (DSR) proposes an upper bound for a test particle's momentum \cite{Amelino2000, Kowalski2005, Amelino2005}. In fact, the existence of a minimal measurable length restricts a test particle's momentum to take a maximal measurable momentum of  the order of the Planck momentum \cite{Magueijo2002, Magueijo2003, Cortes2005}. Several interesting and novel results are obtained  by considering both a minimal length and a maximal momentum \cite{Ali2009, Ali2011*, Nozari2012, Nozari2019}. Moreover, when one considers curvature effects, it can be shown that there is a nonzero minimal uncertainty in momentum measurement too \cite{Hinrichsen1996, Kempf1997}. That is, in large distances, where the curvature of space time becomes important, momentum cannot be precisely determined. With the path integral formulation, such noncommutative background geometries can ultraviolet and infrared regularize quantum field theories in arbitrary dimensions through minimal uncertainties both in positions and in momenta (for more details, see \cite{Hinrichsen1996, Kempf1997}). It is important to note that natural cutoffs are essentially related to the compactness of corresponding symplectic manifold~\cite{Nozari016}. It is well-known that thermodynamic quantities of a black hole can be obtained by the standard uncertainty principle. So, in this respect incorporation of the GUP can modify the black hole physics. Recently, black holes, as a connection between general relativity and quantum mechanics, have been investigated widely in GUP framework. For instance, the GUP prevents black holes from total evaporation. Also the GUP modifies Hawking temperature \cite{Adler1999}. So, because of the existence of a maximal temperature originating from minimal length/maximal energy, the GUP predicts a non-radiating remnant of the order of the Planck mass in the final stage of evaporation. So, while it provide a possible candidate for dark matter\cite{Nozari07012}, it may be also a clue for solving the black hole information loss problem and interestingly opens a realistic door for studying the final stage of black hole evaporation \cite{Adler2001, Camelia2006, Nozari2006*, Kim2008, Banerjee2010, Myung2007, Nozari2012*}. The importance of the subject lies in the fact that black holes are essentially a quantum gravity object and therefore GUP as a phenomenological aspect of quantum gravity provides a more realistic framework to study black hole physics and thermodynamics. This feature has shown its efficiency in recent years study of black hole physics.

With this motivations, we adopt Horndeski gravity with phenomenological quantum gravitational effects to study the black hole thermodynamics. We consider a generalized/extended uncertainty relation that includes a minimal length, a minimal momentum and a maximal momentum to modify the black hole temperature and entropy. Then, by using the modified black hole temperature, we obtain the modified heat capacity. Finally, we consider Parikh-Wilczek tunneling process to describe the Hawking radiation emitted from Horndeski black holes. We compute tunneling rate and also possible correlation between emitted modes. To be more clarified, we study possible correlations between the emitted particles, a feature that can be used by itself to address at least a part of the lost information in the process of black hole formation. The motivation for performing such a study in Horndeski framework is the existence of a gap in this respect in literature in one side, and the fact that Horndeski theory is the most general scalar-tensor theory of gravity where incorporation of quantum gravitational effects may bring new physics in the realm of black hole thermodynamics in this framework.

\section{Horndeski Theory}

In the modern formulation of the Horndeski gravity, the action takes the following form \cite{Babichev2017}
\begin{equation}
S=\int \sqrt{-g}d^4 x \left( \mathcal{L}_2 +\mathcal{L}_3+\mathcal{L}_4+\mathcal{L}_5+\mathcal{L} _4^{bH}+\mathcal{L} _5^{bH} \right) ,
\end{equation}
where $g\equiv det(g_{\mu \nu})$ and $g_{\mu \nu}$ is the metric tensor. In our case, we investigate a class of the Horndeski theory which posses shift symmetry, $\phi \longrightarrow \phi +\textit{constant} $,. It includes six arbitrary functions of the scalar field and its canonical kinetic term which are denoted by $\phi$ and $X=-\partial^{\mu} \phi \partial_{\mu} \phi /2$, respectively. In this notation we consider $G_2, G_3, G_4, G_5 $ for ordinary Horndesky theory and $F_4 , F_5 $ for beyond Horndeski (bH) theory. These are in the following form
\begin{align}
\mathcal{L}_2 &=G_2 ,  \\
\mathcal{L}_3 &=-G_3\Box \phi ,   \\
\mathcal{L}_4 &=G_4 R+G_{4X} [(\Box \phi )^2-(\nabla_{\mu} \nabla_{\nu} \phi )^2], \\
\mathcal{L}_5&=G_5 G_{\mu \nu}\nabla^{\mu} \nabla^{\nu} \phi-\frac{1}{6} G_{5X}[(\Box \phi)^3-3\Box \phi(\nabla_{\mu} \nabla_{\nu} \phi)^2 \nonumber   \\ &+2(\nabla_{\mu} \nabla_{\nu} \phi)^3], \\
\mathcal{L}_4^{bH}&=F_4 \epsilon^{\mu \nu \rho \sigma} \epsilon^{\alpha \beta \gamma }_{\sigma} (\nabla_{\mu} \phi \nabla_{\alpha} \phi) (\nabla_{\nu} \nabla_{\beta} \phi )\nabla_{\rho} \nabla_{\gamma} \phi,  \\
\mathcal{L}_5^{bH}&=F_5 \epsilon^{\mu \nu \rho \sigma} \epsilon^{\alpha \beta \gamma \delta} (\nabla_{\mu} \phi \nabla_{\alpha} \phi) (\nabla_{\nu} \nabla_{\beta} \phi) (\nabla_{\rho} \nabla_{\gamma} \phi )(\nabla_{\sigma} \nabla_{\delta} \phi),
\end{align}
where $R$ is the Ricci scalar, and $G_{\mu \nu}$ is the Einstein tensor. For simplicity, in our notation, $\Box \phi \equiv g^{\mu \nu} \partial_{\mu \nu} \phi$, $(\nabla_{\mu} \nabla_{\nu} \phi)^2 \equiv \nabla_{\mu} \nabla_{\nu} \phi \nabla^{\mu} \nabla^{\nu} \phi $, $(\nabla_{\mu} \nabla_{\nu} \phi)^3 \equiv \nabla_{\mu} \nabla_{\nu} \phi \nabla^{\nu} \nabla^{\rho} \phi \nabla_{\rho} \nabla^{\mu} \phi $ and $G_X= \partial G(X) / \partial X $. Obviously, GR and $f(R)$ gravity are the spatial limits of the Horndeski gravity which are chosen by $G_2=G_3=G_5=0, G_4=1/2$ and $G_2=G_3=G_5=0,  G_4=f(R)$, respectively.

In our case, we are interested in to investigate the spherically symmetric and static black hole solutions in shift symmetric Horndeski theories. These black holes are static and asymptotically flat with a static scalar field \cite{Babichev2017}. So, the static and spherically symmetric ansatz for spacetime and scalar field take the following form respectively:
\begin{equation} \label{eqn8}
ds^2=-f(r)dt^2+\frac{dr^2}{g(r)}+r^2(d\theta^2+sin^2\theta d\varphi^2)  ,
\end{equation}
\begin{equation}
\phi=\phi(r)  .
\end{equation}
Also, we set the $G_i$ functions of the Lagrangian as follows:
\begin{align}
G_2 &=\eta X-2\Lambda
    \nonumber ,  \\
G_4 &=\zeta+\gamma\sqrt{-X},
    \nonumber \\
G_3 &=G_5=F_4=F_5=0,
\end{align}
where $\eta$ and $\gamma$ are dimensionless parameters and $\Lambda$ is the cosmological constant. The first term of $G_2$ is a canonical kinetic term and the first term of $G_4$ is $\zeta=M^2_{Pl}/(16\pi)$ that yields an Einstein-Hilbert term in the action. Finally, the action takes the following form \cite{Babichev2017}
\begin{equation}
S=\int d^4 x \sqrt{-g} \left\{ \left[\zeta+\gamma \sqrt{(\partial\phi)^2/2} \right]R-\frac{\eta}{2} (\partial\phi)^2 -2\Lambda-\frac{\gamma}{\sqrt{2(\partial\phi)^2}} [(\square\phi)^2-(\nabla_{\mu} \nabla_{\nu} \phi)^2] \right\}.
\end{equation}
The scalar field can be obtained from the metric and scalar field ansatz as \cite{Babichev2017}
\begin{equation}
\phi^{'}=\pm \frac{\sqrt{2} \gamma}{\eta r^2 \sqrt{f}}.
\end{equation}
For our particular case, the spacetime metric solution takes the following form \cite{Babichev2017}
\begin{equation} \label{eqn13}
f(r)=g(r)=1- \frac{\mu}{r}- \frac{\gamma^2}{2\zeta \eta r^2}- \frac{\Lambda}{3\zeta}r^2.
\end{equation}
Explicitly, the solution has the Reissner-Nordstr\"{o}m-de sitter (RN+$\Lambda$) form. As a consequence of similarity to the RN+$\Lambda$ form, this solution describes a black hole with mass $\mu/2$ where $\mu$ is a free integration constant and  electric charge $\sqrt{\frac{-\gamma^2}{2\zeta \eta}}$ for spacetime. The parameters $\gamma$ and $\eta$ unavoidably share the same sign. Additionally, this solution has singularity which can be absorbed in the coordinate transformation. Finally, identical to other static solutions, space and time coordinates exchange their roles in the interior of the black hole. Also, $\phi^{'}$ is real for outside of the black hole horizon, $f>0$, and imaginary for interior of the black hole horizon, $f<0$. Further, the solution (\ref{eqn13}) with Eq.(\ref{eqn8}), recovers the Reissner-Nordstr\"{o}m (RN) metric and the Schwarzschild metric in the limits $\Lambda \rightarrow 0$ and $\gamma \rightarrow 0$, respectively. Having introduced a particular black hole solution in shift symmetric Horndeski theory, now we study its thermodynamics in the presence of phenomenological quantum gravitational effects encoded in a GUP relation.

\section{Thermodynamics of Horndeski Black Holes}
To incorporate the GUP effects on the black hole thermodynamics, let us start with the metric to obtain the location of the horizons \cite{Said2011}. The radii of the horizons are determined by the equation $f(r)=0$. In general this equation has four roots, which we can classify them as
\begin{equation}
  r_1>r_2>r_3>r_4
\end{equation}
The lack of cubic term in equation $f(r)=0$ leads to a negative and unphysical root. So, we have three positive (real) roots that the outermost one is the cosmological horizon. $r_2$ and $r_3$ are event horizon and Cauchy horizon. In the current work, we are only interested in the event horizon and  Caushy horizon which are given by
\begin{align}
r_+&=-\frac{1}{2}\sqrt{Y}+\frac{1}{2}\sqrt{\frac{6\zeta}{\Lambda}-Y+\frac{6\zeta\mu}{\Lambda\sqrt{Y}}},
\\ \nonumber \\
r_-&=\frac{1}{2}\sqrt{Y}-\frac{1}{2}\sqrt{\frac{6\zeta}{\Lambda}-Y-\frac{6\zeta\mu}{\Lambda\sqrt{Y}}},
\end{align}
where
\begin{align}
X &= \bigg[ 432\zeta^3 \eta^3 -2592 \gamma^2 \zeta \eta^2 \Lambda-1944 \zeta^2 \eta^3 \Lambda \mu^2 \nonumber \\
 &+\sqrt{ \left[-4 \left(36\zeta^2 \eta^2 +72\gamma^2 \eta \Lambda\right)^3 +\left(432\zeta^3 \eta^3 -2592 \gamma^2 \zeta \eta^2 \Lambda-1944 \zeta^2 \eta^3 \Lambda \mu^2 \right)^2 \right]} \bigg] ^{\frac{1}{3}},
\\   \nonumber  \\
Y &=\frac{2\zeta}{\Lambda}-\frac{6\times 2^{\frac{1}{3}}(\zeta^2 \eta^2+2 \gamma^2 \eta \Lambda)}{\eta\Lambda X}-\frac{X}{6\times 2^{\frac{1}{3}}\eta\Lambda}.
\end{align}
In the standard framework, the uncertainty principle can be used to obtain the temperature and entropy of black hole \cite{Ohanian1994}. So, in the same way the GUP is capable to modify the temperature and entropy. The black hole thermodynamics in the presence of the GUP are investigated extensively \cite{Adler2001, Camelia2006, Nozari2006*, Kim2008, Banerjee2010, Myung2007}. The uncertainty in the position of an emitted particle in the Hawking effect is given by
\begin{equation} \label{eqn19}
\Delta x =2r_+=-\sqrt{Y}+\sqrt{\frac{6\zeta}{\Lambda}-Y
+\frac{6\zeta\mu}{\Lambda\sqrt{Y}}}.
\end{equation}
Also, the uncertainty in the energy of the Hawking particle is
\begin{equation}
\Delta E \approx c \Delta p\approx \frac{\hbar c}{\Delta x}=\hbar c \left[-\sqrt{Y}+\sqrt{\frac{6\zeta}{\Lambda}-Y
+\frac{6\zeta\mu}{\Lambda\sqrt{Y}}}\right]^{-1}.
\end{equation}
In standard cases the black hole temperature is $T_{BH}= \frac{\kappa}{2\pi}=T_0$, where $\kappa$ and $T_0$ are the horizon surface gravity and the Hawking temperature, respectively. What is the black hole temperature in Horndeski theory? For our special case, considering black hole solution Eq.(\ref{eqn13}) yields the standard formula again and there is not any temperature shift (for more details see \cite{Hajian2021}). So, the Hawking temperature is associated to the black hole event  horizon radius by
\begin{equation} \label{eqn21}
T=\frac{1}{4\pi r_+}=\frac{1}{2 \pi \Delta x}\,.
\end{equation}
Using the Hawking temperature, the Bekenstein-Hawking entropy related to the black hole mass has the following well known form
\begin{equation} \label{eqn22}
T=\frac{dE}{dS}=\frac{dM}{dS}\,.
\end{equation}
Using Eq.(\ref{eqn19}), Eq.(\ref{eqn21}) and Eq.(\ref{eqn22}), we find
\begin{equation} \label{eqn23}
S= 2 A \left(-\sqrt{Y}+\sqrt{\frac{6\zeta}{\Lambda}-Y
+\frac{6\zeta\mu}{\Lambda\sqrt{Y}}} \right)^{-2}\times \int dM \left[-\sqrt{Y}+\sqrt{\frac{6\zeta}{\Lambda}-Y
+\frac{6\zeta\mu}{\Lambda\sqrt{Y}}} \right]\,,
\end{equation}
where $A=4\pi r_+^{2}$ is considered as the surface area of the black hole event horizon. Note that the solution (\ref{eqn23}) recovers $T_{Sch}= 4 \pi M^2$ in appropriate limit where $T_{Sch}$ is the entropy of the Schwarzschild black hole in the standard framework. To incorporate the quantum gravity effects on the black hole thermodynamics in this shift symmetric Horndeski theory, we take into account the GUP and develop the study in more details. We consider a general uncertainty principle that includes a minimal length, a minimal momentum and a maximal momentum. This GUP has the following form \cite{Kempf1994, Kempf1995, Hinrichsen1996, Kempf1997}
\begin{equation} \label{eqn24}
\Delta x \Delta p \geq \hbar \Big[1-\alpha \ell_p (\Delta p) + \alpha^2 \ell_p^{2} (\Delta p)^2+ \beta^2 \ell_p^{2} (\Delta x)^2\Big],
\end{equation}
where $\ell_p$ is Planck length, $\alpha$ and $\beta$ are dimensionless parameters which normally are of the order of unity and depend on the quantum gravity approaches. Solving this relation for $\Delta p$ gives us the following momentum uncertainty
\begin{equation}
\frac{\Delta p}{\hbar} = \frac{(\alpha \hbar \ell_p+2r_+)}{2\alpha^2 \hbar^2 \ell_p^{2}} \left(1\pm\sqrt{1-\frac{4\alpha^2 \hbar \ell_p^{2}(\hbar+4\beta^2 \hbar \ell_p^{2} r_+^{2})}{(\alpha \hbar \ell_p+2r_+)^2}} \right).
\end{equation}
We can show that this solution has a minimal length, $(\Delta x)_{min}=\alpha \ell_p$, , a minimal momentum, $(\Delta p)_{min}=2 \beta \ell_p$, and a maximal momentum, $(\Delta p)_{max}\simeq \frac{1}{\alpha \ell_p}$ with $c=1$. Using the series expansion
\begin{align} \label{eqn26}
\frac{\Delta p}{\hbar}&= \frac{1}{\Delta x} \bigg( 1-\frac{\alpha\hbar \ell_p}{\Delta x}+\frac{(2\alpha^2 \hbar^2 +\beta^2 \Delta x^4) \ell_p^{2}}{\Delta x^2}- \frac{\alpha \hbar (4 \alpha^2 \hbar^2+ \beta^2 \Delta x^4)\ell_p^{3}}{\Delta x^3}   \nonumber \\
&+ \frac{3 \alpha^2 \hbar^2 (3 \alpha^2 \hbar^2+ \beta^2 \Delta x^4) \ell_p^{4}}{\Delta x^4}+O(\ell_p^{5}) \bigg)\,,
\end{align}
and substituting Eq.\eqref{eqn26} into Eq.\eqref{eqn24}, we find the GUP corrected position uncertainty as
\begin{align} \label{eqn27}
\Delta x^{'} &= \Delta x \Bigg[ \Bigg(1-\frac{\alpha\hbar \ell_p}{\Delta x}+\frac{(2\alpha^2 \hbar^2+\beta^2 \Delta x^4) \ell_p^{2}}{\Delta x^2}-\frac{\alpha \hbar (4 \alpha^2 \hbar^2+ \beta^2 \Delta x^4)\ell_p^{3}}{\Delta x^3} \nonumber   \\
 &+\frac{3 \alpha^2 \hbar^2 (3 \alpha^2 \hbar^2+ \beta^2 \Delta x^4) \ell_p^{4}}{\Delta x^4} \Bigg)^{-1}-\frac{\alpha \hbar \ell_p}{\Delta x} +\frac{\alpha^2 \ell_p^{2} \hbar^2}{\Delta x^2} \Bigg(1-\frac{\alpha\hbar \ell_p}{\Delta x}+\frac{(2\alpha^2 \hbar^2 +\beta^2 \Delta x^4) \ell_p^{2}}{\Delta x^2} \nonumber \\
 &-\frac{\alpha \hbar(4 \alpha^2 \hbar^2+\beta^2 \Delta x^4)\ell_p^{3}}{\Delta x^3}
+ \frac{3 \alpha^2 \hbar^2 (3 \alpha^2 \hbar^2+ \beta^2 \Delta x^4) \ell_p^{4}}{\Delta x^4} \Bigg)+\beta^2 \ell_p^{2} \Delta x^2 \Bigg(1-\frac{\alpha\hbar \ell_p}{\Delta x}  \nonumber  \\
&+\frac{(2\alpha^2 \hbar^2 +\beta^2 \Delta x^4) \ell_p^{2}}{\Delta x^2}- \frac{\alpha \hbar (4 \alpha^2 \hbar^2+\beta^2 \Delta x^4)\ell_p^{3}}{\Delta x^3}+ \frac{3 \alpha^2 \hbar^2 (3 \alpha^2 \hbar^2+ \beta^2 \Delta x^4) \ell_p^{4}}{\Delta x^4} \Bigg)^{-3}    \nonumber   \\
 &-2\hbar \alpha \beta^2 \ell_p^{3} \Delta x \Bigg(1-\frac{\alpha\hbar \ell_p}{\Delta x}+\frac{(2\alpha^2 \hbar^2+\beta^2 \Delta x^4) \ell_p^{2}}{\Delta x^2}-\frac{\alpha \hbar (4 \alpha^2 \hbar^2+ \beta^2 \Delta x^4)\ell_p^{3}}{\Delta x^3} \nonumber  \\
 &+\frac{3 \alpha^2 \hbar^2 (3 \alpha^2 \hbar^2+ \beta^2 \Delta x^4) \ell_p^{4}}{\Delta x^4} \Bigg)^{-2} \Bigg].
\end{align}
So, the modified Hawking temperature  for a GUP-corrected Horndeski black hole without charge can be obtained as follows
\begin{align} \label{eqn28}
T^{'} &=\frac{1}{2\pi \Delta x^{'}}=T \Bigg[ \Bigg(1-\frac{\alpha\hbar \ell_p}{\Delta x}+\frac{(2\alpha^2 \hbar^2+\beta^2 \Delta x^4) \ell_p^{2}}{\Delta x^2}-\frac{\alpha \hbar (4 \alpha^2 \hbar^2+ \beta^2 \Delta x^4)\ell_p^{3}}{\Delta x^3}  \nonumber \\
  &+\frac{3 \alpha^2 \hbar^2 (3 \alpha^2 \hbar^2+ \beta^2 \Delta x^4) \ell_p^{4}}{\Delta x^4} \Bigg)^{-1}-\frac{\alpha \hbar \ell_p}{\Delta x} +\frac{\alpha^2 \ell_p^{2} \hbar^2}{\Delta x^2} \Bigg(1-\frac{\alpha\hbar \ell_p}{\Delta x} + \frac{(2\alpha^2 \hbar^2 + \beta^2 \Delta x^4) \ell_p^{2}}{\Delta x^2} \nonumber \\
 &-\frac{\alpha \hbar(4 \alpha^2 \hbar^2+\beta^2 \Delta x^4)\ell_p^{3}}{\Delta x^3}   +\frac{3 \alpha^2 \hbar^2 (3 \alpha^2 \hbar^2+ \beta^2 \Delta x^4)\ell_p^{4}}{\Delta x^4} \Bigg)+\beta^2 \ell_p^{2} \Delta x^2 \Bigg(1-\frac{\alpha\hbar \ell_p}{\Delta x} \nonumber \\
 &+\frac{(2\alpha^2 \hbar^2 +\beta^2 \Delta x^4) \ell_p^{2}}{\Delta x^2}- \frac{\alpha \hbar (4 \alpha^2 \hbar^2+\beta^2 \Delta x^4)\ell_p^{3}}{\Delta x^3}+ \frac{3 \alpha^2 \hbar^2 (3 \alpha^2 \hbar^2+ \beta^2 \Delta x^4) \ell_p^{4}}{\Delta x^4} \Bigg)^{-3}   \nonumber  \\
&-2\hbar \alpha \beta^2 \ell_p^{3} \Delta x \Bigg( 1-\frac{\alpha\hbar \ell_p}{\Delta x}    +\frac{(2\alpha^2 \hbar^2+\beta^2 \Delta x^4) \ell_p^{2}}{\Delta x^2}-\frac{\alpha \hbar (4 \alpha^2 \hbar^2+ \beta^2 \Delta x^4)\ell_p^{3}}{\Delta x^3} \nonumber   \\
 &+\frac{3 \alpha^2 \hbar^2 (3 \alpha^2 \hbar^2+ \beta^2 \Delta x^4) \ell_p^{4}}{\Delta x^4} \Bigg)^{-2} \Bigg]^{-1}\,,
\end{align}
where $\Delta x$ is given by Eq.\eqref{eqn19} and $T$ is the standard Bekenstein-Hawking temperature. This allows us to calculate the modified entropy from Eq.\eqref{eqn22} as follows
\begin{equation} \label{eqn29}
S^{'}(M)=S(M)-7 \ell_p^{5} F(M)+O(\ell_p^{6})  ,
\end{equation}
where $F(M)$ is given by
\begin{align} \label{eqn30}
F(M)&=\int dM \Bigg[2 A \Big( \left( \alpha^3 \beta^2 \hbar^3 \right) \left(-\sqrt{Y}+\sqrt{\frac{6\zeta}{\Lambda}-Y
+\frac{6\zeta\mu}{\Lambda\sqrt{Y}}} \right)^{-2}  \nonumber     \\
&+ \left(3\alpha^5 \hbar^5 \right) \left( -\sqrt{Y}+\sqrt{\frac{6\zeta}{\Lambda}-Y
+\frac{6\zeta\mu}{\Lambda\sqrt{Y}}} \right)^{-6} \Big) \Bigg]\,,
\end{align}
and the surface area of the black hole's outer horizon is given by
\begin{equation}
 A=4\pi r_+^{2}\,.
\end{equation}
We note that the existence of maximal momentum and minimal momentum leads to extra terms in Hawking temperature and we have both of even and odd powers of Planck length, $\ell_p$, in comparison to the results reported in \cite{Said2011}. Further, the $F(M)$ term in the (\ref{eqn30}) only exists in the presence of the GUP. So, in the absence of GUP, this term vanishes and Eq.(\ref{eqn30}) reduces to the Schwarzschild entropy in the standard framework as is expected.

\subsection{Heat Capacity}

The heat capacity of black hole, in the semiclassical approach, can be obtained by the inverse temperature, $T^{-1}=\beta= \frac{dS}{dM}$. Generally, the heat capacity can be calculated as follows
\begin{equation}
  C=\frac{dM}{dT}
\end{equation}
\begin{equation} \label{eqn33}
  C=\frac{1}{\pi T^2}\left[\frac{1}{\sqrt{Y}}+\left(\frac{6\zeta}{\Lambda}-Y+\frac{12\zeta M}{\Lambda \sqrt{Y}} \right)^{-\frac{1}{2}} \left(1+\frac{6\zeta M}{\Lambda Y^{\frac{3}{2}}} \right)\right]^{-1} \frac{1}{W},
\end{equation}
where W and Z are given by
\begin{equation}
  W= \frac{dY}{dX}Z = \left[\frac{6\times 2^{\frac{1}{3}} \left(\zeta^2 \eta^2 +2 \gamma^2 \eta \Lambda \right)}{\eta \Lambda X^2}- \frac{1}{6\times 2^{\frac{1}{3}}\eta \Lambda} \right]Z
\end{equation}
and
\begin{align}
  Z&= \frac{dX}{dM}= \frac{1}{3} X^{-2} \bigg(15552 \zeta^2 \eta^3 \Lambda M+\frac{1}{2} \Big[ X^3 -432 \zeta^3 \eta^3 -592 \gamma^2 \zeta \eta^2 \Lambda \nonumber \\
   & -7774 \zeta^2 \eta^3 \Lambda M \Big]^{-1} \left[2 \left(432 \zeta^3 \eta^3-2592 \gamma^2 \zeta \eta^2 \Lambda -7776 \zeta^2 \eta^3 \Lambda M \right)15552 \zeta^2 \eta^3 \Lambda M \right] \bigg).
\end{align}
If we consider the GUP, we get to
\begin{align} \label{eqn36}
 C^{'}&=C \left[1-7\ell_p^{5}
 \left(\frac{\alpha^3 \beta^2 \hbar^3}{(-\sqrt{Y}+\sqrt{\frac{6\zeta}{\Lambda}-Y
 +\frac{6\zeta\mu}{\Lambda\sqrt{Y}}})}+\frac{3\alpha^5 \hbar^5}{(-\sqrt{Y}+\sqrt{\frac{6\zeta}{\Lambda}-Y
 +\frac{6\zeta\mu}{\Lambda\sqrt{Y}}})^5} \right) \right]^2   \nonumber  \\
 &\left[7\ell_p^{5}
 \left(\frac{\alpha^3 \beta^2 \hbar^3}{\left(-\sqrt{Y}+\sqrt{\frac{6\zeta}{\Lambda}-Y
 +\frac{6\zeta\mu}{\Lambda\sqrt{Y}}}\right)^2}+\frac{15\alpha^5 \hbar^5}{\left(-\sqrt{Y}+\sqrt{\frac{6\zeta}{\Lambda}-Y
 +\frac{6\zeta\mu}{\Lambda\sqrt{Y}}} \right)^6} \right) \right]^{-1}\,,
\end{align}
where $C$ is the standard Bekenstein-Hawking heat capacity.  When $\alpha$ and $\gamma$, the GUP parameters tend to zero, Eq.\eqref{eqn36} reduces to Eq.\eqref{eqn33} as would be expected.

\subsection{Tunneling Process}
In 1974, Stephen Hawking demonstrated \cite{Hawking1974} that black holes have an emission spectrum of a black body, the so called Hawking radiation, and so are not purely black. In 2000, Parikh and Wilczek exhibited \cite{Parikh2000} a semiclassical method to derive Hawking radiation as a tunneling process from the event horizon of black hole. In this section we calculate Hawking radiation in Parikh-Wilczek tunneling formalism. The coordinate system should be well-behaved for calculations at the event horizon. So, we define the Painlev\'{e}-like coordinate transformation as follows \cite{Medved2002,Miao2011, Qing2006}

\begin{equation} \label{eqn37}
  dt_R=dt+f^{'}(r) dr  ,
\end{equation}
where $t_R$ is the black hole time coordinate. Substituting Eq.\eqref{eqn37} into Eq.\eqref{eqn8} we have
\begin{align}
  ds^2 &=-\Delta dt_R^2 + \frac{1}{\Delta} dr^2 +r^2 d\Omega^2
  \nonumber   \\
  &=-\Delta dt^2 + (-\Delta f^{'2}+\frac{1}{\Delta})dr^2 -2\Delta f^{'}drdt+r^2d\Omega^2\,,
\end{align}
where simplicity we have defined $\Delta=1-\frac{\mu}{r}-\frac{\gamma^2}{2\zeta\eta r^2}-\frac{\Lambda}{3\zeta}r^2$. Then $f^{'} (r)$ satisfies
\begin{equation}
  f^{'}=\pm \frac{\sqrt{1-\Delta}}{\Delta}
\end{equation}
and the Painlev\'{e} line element and the radial geodesics take the following form respectively
\begin{equation}
  ds^2=-\Delta dt^2+dr^2\mp 2\sqrt {1-\Delta} drdt+r^2d \Omega\,,
\end{equation}
\begin{equation}
  \dot{r}=\frac{dr}{dt}=\pm 1\mp\sqrt{\frac{\mu}{r}+\frac{\gamma^2}{2\zeta\eta r^2}+\frac{\Lambda}{3\zeta}r^2}.
\end{equation}
In this process, that occurs near inside the horizon, the particle with positive energy, $\tilde{w}$,  tunnels out and escapes the event horizon. Considering the energy conservation, the mass parameter will be replaced with $\mu \rightarrow \mu - \tilde{w}$. We can rewrite the new line element and the radial null geodesics which are respectively as
\begin{equation}
 ds^{2}=-\tilde{\Delta} dt^2+dr^2\mp 2\sqrt {1-\tilde{\Delta}} drdt+r^2d \Omega\,,
\end{equation}
and
\begin{equation}
  \dot{r}=\frac{dr}{dt}=\pm 1\mp\sqrt{\frac{(\mu-\tilde{\omega})}{r}+\frac{\gamma^2}{2\zeta\eta r^2}+\frac{\Lambda}{3\zeta}r^2} ,
\end{equation}
where $ \tilde{\Delta}=1-\frac{(\mu-\tilde{\omega})}{r}-\frac{\gamma^2}{2\zeta\eta r^2}-\frac{\Lambda}{3\zeta}r^2$.
To compute the tunneling rate, as a semi-classical procedure, we consider the Wentzel-Kramers-Brillouin (WKB) approximation. The tunneling probability is the imaginary part of the action
\begin{equation} \label{eqn44}
  \Gamma \sim \exp (-2\, \textrm{Im}\, \textit{S}) .
\end{equation}
The imaginary part of the particle action across the event horizon, $r_+$, from initial position, $r_{in}$, to the final position, $r_{out}$, is defined as
\begin{equation} \label{eqn45}
\textrm{Im}\, \textit{S}=\textrm{Im}\, \int_{r_{in}}^{r_{out}} p_r dr= \textrm{Im}\, \int_{r_{in}}^{r_{out}} \int_{0}^{p_r} d{\tilde{p}_r} dr,
\end{equation}
where $p_r$ is the canonical momentum of the outgoing particle. By using the Hamilton's canonical equation
\begin{equation} \label{eqn46}
 \dot{r}=\frac{dH}{dp_r}= \frac{d(\mu - \tilde{\omega})}{dp_r}\,,
\end{equation}
and by substituting Eq.\eqref{eqn46} into Eq.\eqref{eqn45} we get
\begin{equation}
  \textrm{Im}\, S = \textrm{Im}\, \int_{r_{in} }^{r_{out}} \int_{0}^{\omega} \frac{(-d\tilde{\omega}) dr}{\dot{r}}=Im \int_{r_{in}}^{r_{out}} \int_{0}^{\omega}\frac{(-d\tilde{\omega}) dr}{1-\sqrt{\frac{(\mu-\tilde{\omega})}{r}+\frac{\gamma^2}{2\zeta\eta r^2}+\frac{\Lambda}{3\zeta}r^2}}.
\end{equation}
The commutation relation, from the GUP expression, in the presence of minimal length, minimal momentum and maximal momentum is
\begin{equation}
  \left[r,p_r \right]=i \left(1-\alpha \ell_p p +\alpha^2 \ell_p^2 p^2+\beta^2 \ell_p^2 r^2 \right).
\end{equation}
In the classical limit we can rewrite this relation between the radial coordinate and the conjugate momentum by poisson bracket
\begin{equation}
\{r,p_r\}= \left(1-\alpha \ell_p p +\alpha^2 \ell_p^2 p^2+\beta^2 \ell_p^2 r^2 \right).
\end{equation}
So, we obtain the deformed Hamiltonian equation as follows
\begin{equation}
  \dot{r}=\{r,H\} =\{r,p_r\} \frac{dH}{dr} .
\end{equation}
Finally, we can rewrite the imaginary part of the action in the presence of GUP as follows
\begin{align}
\textrm{ Im }\, S &= \textrm{Im}\, \int_{r_{in}}^{r_{out}} \int_{0}^{\omega} \frac{\hbar(1-\alpha \ell_p \tilde{\omega} +\alpha^2 \ell_p^2 \tilde{\omega}^{2} )}{1-\sqrt{\frac{(\mu-\tilde{\omega})}{r}+\frac{\gamma^2}{2\zeta\eta r^2}+ \frac{\Lambda}{3\zeta}r^2}} (-d\tilde{\omega}) dr \nonumber    \\
 &+\textrm{Im}\, \int_{r_{in}}^{r_{out}} \int_{0}^{\omega} \frac{\hbar(\gamma^2 \ell_p^2r^2)}{1-\sqrt{\frac{(\mu-\tilde{\omega})}{r}+\frac{\gamma^2}{2\zeta\eta r^2}+\frac{\Lambda}{3\zeta}r^2}} (-d\tilde{\omega}) dr .
\end{align}
The integral takes the following form
\begin{align}
\textrm{Im}\,S &= \textrm{Im}\, \int_{r_{in}}^{r_{out}} \bigg[2\pi r -2 \alpha \ell_p \pi \left(\mu r - r^2 + Q^2+ \frac{\Lambda r^4}{3\zeta}\right)+2\pi \alpha^2 \ell^2 \Big[ \left(\mu^2 -2 Q^2 \right)r
  -2\mu r^2      \nonumber   \\
 &+2\mu Q^2 +\frac{2\mu \Lambda r^4}{3\zeta}+\left(1+\frac{2 Q^2 \Lambda}{3\zeta} \right) r^3 +Q^4 \frac{1}{r} +\left(\frac{\Lambda}{3\zeta} \right)^2 r^7 +\frac{2\Lambda r^5}{3\zeta} \Big]+2 \pi\beta^2 \ell_p^2 r^3 \bigg]dr\,,
\end{align}
where $Q=\frac{\gamma^2}{2\zeta\eta}$. Therefore, the imaginary part of the action takes the following form
\begin{align} \label{eqn53}
 \textrm{Im}\, \textit{S} &=-\pi \left(r_{out}^2-r_{in}^2\right) \left[-1+\alpha \ell_p \mu -\alpha^2 \ell_p^{2} \mu^2+2\alpha^2 \ell_p^{2} Q^2\right]-2\pi \frac{\left(r_{out}^3-r_{in}^3\right)}{3}\left[-\alpha \ell_p+2\alpha^2 \ell_p^{2}\mu \right]
  \nonumber   \\
   &- 2\pi \left(r_{out}-r_{in}\right) \left[2\alpha \ell_p Q^2-2\alpha^2 \ell_p^{2}\mu Q^2\right]-2\pi \frac{\left(r_{out}^5-r_{in}^5\right)}{5}\left[\alpha \ell_p \frac{\Lambda}{3\zeta}-2\mu \Lambda \alpha^2 \ell_p^{2}\right] \nonumber      \\
  &+2 \pi \alpha^2 \ell_p^{2} \frac{\Lambda}{9\zeta} \left(r_{out}^6-r_{in}^6\right)
  -\pi \frac{\left(r_{out}^4-r_{in}^4\right)}{2}\left[\beta^2 \ell_p^{2}-2\alpha^2 \ell_p^{2} \frac{Q^2 \Lambda}{3\zeta}\right]+2\pi \alpha^2 \ell_p^{2} Q^4 \left(\ln r_{out}-\ln r_{in}\right)
    \nonumber   \\
  &+2\pi \alpha^2 \ell_p^{2}\left(\frac{\Lambda}{3\zeta}\right)^2 \frac{\left(r_{out}^8-r_{in}^8 \right)}{8}   .
\end{align}
Substituting Eq.\eqref{eqn53} into Eq.\eqref{eqn44}, we obtain the tunneling rate at the horizon as follows
\begin{align}
\Gamma &\approx \exp \Bigg\{2 \pi \left(r_{out}^2-r_{in}^2\right) \left[-1+\alpha \ell_p \mu -\alpha^2 \ell_p^{2} \mu^2+2\alpha^2 \ell_p^{2} Q^2 \right]+4 \pi \frac{\left(r_{out}^3-r_{in}^3\right)}{3}\left[-\alpha \ell_p+2\alpha^2 \ell_p^{2}\mu \right]
  \nonumber   \\
   &+4 \pi \left(r_{out}-r_{in} \right) \left[2\alpha \ell_p Q^2-2\alpha^2 \ell_p^{2}\mu Q^2 \right]+4 \pi \frac{\left(r_{out}^5-r_{in}^5 \right)}{5}\left[\alpha \ell_p \frac{\Lambda}{3\zeta}-2\mu \Lambda \alpha^2 \ell_p^{2} \right] \nonumber      \\
  &-4 \pi \alpha^2 \ell_p^{2}\frac{\Lambda}{9\zeta} \left(r_{out}^6-r_{in}^6 \right)+2\pi \frac{\left(r_{out}^4-r_{in}^4 \right)}{2}\left[\beta^2 \ell_p^{2}-2\alpha^2 \ell_p^{2} \frac{Q^2 \Lambda}{3\zeta} \right]-4 \pi \alpha^2 \ell_p^{2} Q^4 \left(\ln r_{out}-\ln r_{in} \right)
    \nonumber   \\
  &-4 \pi \alpha^2 \ell_p^{2} \left(\frac{\Lambda}{3 \zeta} \right)^2 \frac{\left(r_{out}^8-r_{in}^8\right)}{8} \Bigg\}= \exp \left(\Delta \textit{S}_{BH} \right),
\end{align}
where $\Delta \textit{S}_{BH}=\textit{S}_{BH} (\mu -\omega)-\textit{S}_{BH}(\mu) $ is the difference in Bekenstein-Hawking entropy before and after the particles emission at the event horizon. When $\gamma=0$ and $\Lambda=0$, the result reduces to the Schwarzschild black hole's result \cite{Parikh2000,Medved2002}. Because of the extra terms in comparison to the results of \cite{Nozari2012*}, the emission spectrum is not purely thermal.

Finally, we calculate the possible correlation between the emitted particles (modes) that can be obtained by the following relation

\begin{equation}\label{55}
  \chi(E_1+E_2; E_1,E_2) \equiv \ln [\Gamma (E_1+E_2)]- \ln [\Gamma (E_1) \Gamma(E_2)]\,,
\end{equation}
where $\ln [\Gamma (E_1)]$ and $\ln[\Gamma (E_2)]$ are the emission rates for the first and second emitted particles and $\ln [\Gamma (E_1+E_2)]$ is the emission rate for a single, composed particle with energy $E= E_1+E_2$. The emission rate for the first quanta that carries out the energy $E_1$ is given by

\begin{align}
\ln [\Gamma (E_1)] &=2 \pi r^2 \left[-1+\alpha \ell_p (\mu-E_1) -\alpha^2 \ell_p^{2} (\mu-E_1)^2+2\alpha^2 \ell_p^{2} Q^2 \right]+ \frac{4 \pi r^3}{3}\left[-\alpha \ell_p+2\alpha^2 \ell_p^{2}(\mu-E_1) \right]
  \nonumber   \\
   &+4 \pi r \left[2\alpha \ell_p Q^2-2\alpha^2 \ell_p^{2}(\mu-E_1) Q^2 \right]+ \frac{4 \pi r^5}{5}\left[\alpha \ell_p \frac{\Lambda}{3\zeta}-2(\mu-E_1) \Lambda \alpha^2 \ell_p^{2} \right] \nonumber      \\
  &-4 \pi \alpha^2 \ell_p^{2}\frac{\Lambda}{9\zeta} r^6+ \frac{2\pi r^4}{2}\left[\beta^2 \ell_p^{2}-2\alpha^2 \ell_p^{2} \frac{Q^2 \Lambda}{3\zeta} \right]-4 \pi \alpha^2 \ell_p^{2} Q^4 \ln r
  -\frac{4 \pi \alpha^2 \ell_p^{2}r^8}{8} \left(\frac{\Lambda}{3 \zeta} \right)^2.
\end{align}
Similarly, the emission rate for the second quanta that carries out the energy $E_2$ is as follows
\begin{align}
\ln [\Gamma (E_2)] &=2 \pi r^2 \left[-1+\alpha \ell_p ((\mu-E_1)-E_2) -\alpha^2 \ell_p^{2} ((\mu-E_1)-E_2)^2+2\alpha^2 \ell_p^{2} Q^2 \right]+
  \nonumber   \\
   &\frac{4 \pi r^3}{3}\left[-\alpha \ell_p+2\alpha^2 \ell_p^{2}((\mu-E_1)-E_2) \right]+4 \pi r \left[2\alpha \ell_p Q^2-2\alpha^2 \ell_p^{2}((\mu-E_1)-E_2) Q^2 \right]+ \nonumber      \\
  &\frac{4 \pi r^5}{5}\left[\alpha \ell_p \frac{\Lambda}{3\zeta}-2((\mu-E_1)-E_2) \Lambda \alpha^2 \ell_p^{2} \right] -4 \pi \alpha^2 \ell_p^{2}\frac{\Lambda}{9\zeta} r^6+ \frac{2\pi r^4}{2}\left[\beta^2 \ell_p^{2}-2\alpha^2 \ell_p^{2} \frac{Q^2 \Lambda}{3\zeta} \right]
  \nonumber   \\
  &-4 \pi \alpha^2 \ell_p^{2} Q^4 \ln r -\frac{4 \pi \alpha^2 \ell_p^{2}r^8}{8} \left(\frac{\Lambda}{3 \zeta} \right)^2.
\end{align}
Now, the emission rate for a single quanta that carries out the energy $E_1+E_2$ is given by
\begin{align}
\ln [\Gamma (E_1+E_2)] &=2 \pi r^2 \left[-1+\alpha \ell_p (\mu-E_1-E_2) -\alpha^2 \ell_p^{2} (\mu-E_1-E_2)^2+2\alpha^2 \ell_p^{2} Q^2 \right]+
  \nonumber   \\
   &\frac{4 \pi r^3}{3}\left[-\alpha \ell_p+2\alpha^2 \ell_p^{2}(\mu-E_1-E_2) \right]+4 \pi r \left[2\alpha \ell_p Q^2-2\alpha^2 \ell_p^{2}(\mu-E_1-E_2) Q^2 \right]+ \nonumber      \\
  &\frac{4 \pi r^5}{5}\left[\alpha \ell_p \frac{\Lambda}{3\zeta}-2(\mu-E_1-E_2) \Lambda \alpha^2 \ell_p^{2} \right]-4\pi \alpha^2 \ell_p^{2}\frac{\Lambda}{9\zeta} r^6+ \frac{2\pi r^4}{2}\left[\beta^2 \ell_p^{2}-2\alpha^2 \ell_p^{2} \frac{Q^2 \Lambda}{3\zeta} \right]
  \nonumber   \\
  &-4 \pi \alpha^2 \ell_p^{2} Q^4 \ln r -\frac{4 \pi \alpha^2 \ell_p^{2}r^8}{8} \left(\frac{\Lambda}{3 \zeta} \right)^2.
\end{align}
The non-zero statistical correlation function can be calculated as
\begin{align}\label{56}
 \chi (E_1+E_2; E_1,E_2) &=(-2\pi r^2 -4\pi^2 r^4)+\Big(8\pi Q^2 r \alpha- \frac{4}{3} \pi r^3 \alpha+ \frac{4\pi r^5 \alpha \Lambda}{15\zeta}+2\pi r^2 \alpha (\mu -E_1 -E_2)
 \nonumber   \\
 &+\frac{4\pi^2 r^3 (120 Q^2 \alpha \zeta-20 r^2 \alpha \zeta +4 r^4 \alpha \Lambda +30 r \alpha \zeta \mu-30r \alpha \zeta E_1 -15 r \alpha \zeta E_2)}{15\zeta} \Big)\ell_p
    \nonumber   \\
 &+(-4 \ln r \pi Q^4 \alpha^2 - \frac{4\pi r^6 \alpha^2 \Lambda}{9\zeta}-\frac{\pi r^8 \alpha^2 \Lambda^2}{18\zeta^2}+\pi r^4(\beta^2-\frac{2 Q^2 \alpha^2 \Lambda}{3\zeta})
 \nonumber   \\
 &+2\pi r^2(-4 \ln r \pi Q^4 \alpha^2 -\frac{4\pi r^6 \alpha^2 \Lambda}{9\zeta}-\frac{\pi r^8 \alpha^2 \Lambda^2}{18\zeta^2}+\pi r^4(\beta^2-\frac{2 Q^2 \alpha^2 \Lambda}{3\zeta})
 \nonumber  \\
 &+2\pi r^2(2Q^2 \alpha^2-\alpha^2 (\mu -E_1)^2)-\frac{8}{5} \pi r^5 \alpha^2 \Lambda (\mu-E_1)+\frac{8}{3}\pi r^3 \alpha^2 (\mu-E_1)
  \nonumber \\
   &+8 \pi Q^2 r \alpha^2 (-\mu+E_1)) \Big(8 \pi Q^2 r \alpha -\frac{4}{3} \pi r^3 \alpha +\frac{4 \pi r^5 \alpha \Lambda}{15 \zeta}+2\pi r^2 \alpha(\mu-E_1) \Big)
   \nonumber  \\
   & \Big(8 \pi Q^2 r \alpha -\frac{4}{3} \pi r^3 \alpha +\frac{4 \pi r^5 \alpha \Lambda}{15 \zeta}+2\pi r^2 \alpha(\mu-E_1-E_2) \Big)-\frac{8}{5}\pi r^5 \alpha^2 \Lambda (\mu-E_1-E_2)
    \nonumber  \\
  &+\frac{8}{3}\pi r^3 \alpha^2 (\mu-E_1-E_2)+8 \pi Q^2 r \alpha^2 (-\mu+E_1+E_2)+2 \pi r^2 \alpha^2 (2Q^2-(\mu-E_1-E_2)^2)
  \nonumber  \\
  &+2\pi r^2 \Big(-4 \ln r \pi Q^4 \alpha^2 -\frac{4\pi r^6 \alpha^2 \Lambda}{9\zeta}-\frac{\pi r^8 \alpha^2 \Lambda^2}{18 \zeta^2}+\pi r^4 \big(\beta^2 -\frac{2 Q^2 \alpha^2 \Lambda}{3\zeta}
   \nonumber   \\
   &-\frac{8}{5}\pi r^5 \alpha^2 \Lambda(\mu-E_1-E_2)+\frac{8}{3}\pi r^3 \alpha^2 (\mu-E_1-E_2)+8 \pi Q^2 r \alpha^2(-\mu+E_1+E_2)
  \nonumber  \\
  & +2 \pi r^2 \alpha^2(2 Q^2 -(\mu-E_1-E_2)^2) \big)\Big)\ell_p^{2}+O(\ell_p^{3}) .
\end{align}
 Obviously, the statistical correlation function is not zero. So, black hole radiation is not purely thermal. Also, existence of the non-zero correlation means that information can come out during the evaporation process. Since these correlations can store some sort of information, so these correlations are capable to address at least part of the lost information in essence.

\section{Conclusion}
While the issue of black hole thermodynamics in the framework of the generalized/extended uncertainty relations has been studied vastly in literature, the issue of Horndeski black holes' thermodynamics in the framework of GUP/EUP has been overlooked in literature. On the other hand, Horndeski theory provides the most general framework for scalar-tensor theories of gravity. For these reasons, in this paper we have focused on the thermodynamics of shift symmetric Horndeski black hole solutions in the framework of phenomenological quantum gravity corrections encoded in a class of generalized/extended uncertainty relation. We obtained in details the temperature and then the heat capacity of such a black hole that recovers the standard Schwarzschild, Reissner-Nordstr\"{o}m or Reissner-Nordstr\"{o}m-de sitter solutions in the appropriate limits. Then the issue of Hawking radiation as a semi-classical tunneling from the event horizon has been studied in details. For this purpose, the imaginary part of the classical action has been calculated within the WKB approximation. The issue of possible correlations between the emitted modes (particles) has been treated carefully and it is shown that these correlations are not vanishing, leading to the conclusion that part of the lost information may be stored in these quantum gravitational correlations. It is important to note that we focused mainly on the near, ``event" horizon calculations. There is in fact some correlations between the various horizons of these multi-horizon geometry \cite{Shanki2003} and these correlations should be taken into account in a more realistic and concrete study. We leave this issue for our forthcoming study.

\end{document}